\documentclass[useAMS]{mn2e}
\usepackage{times}
\usepackage{amssymb}
\usepackage{rotating}
\input{epsf}

\title[Power spectra of atolls]
{A physical interpretation of the variability power spectral
components in accreting neutron stars}

\author[A. Ingram \& Chris Done]
{Adam
Ingram$^{1}\thanks{E-mail:a.r.ingram@durham.ac.uk}$ \&
Chris Done$^{1}$\\
$^1$Department of Physics, University of Durham, South Road,
Durham DH1 3LE, UK\\
}

\date{Submitted to MNRAS}
\pagerange{\pageref{firstpage}--\pageref{lastpage}} \pubyear{2009}

\begin{document}

\topmargin = -0.5cm

\maketitle

\label{firstpage}

\begin{abstract}

We propose a physical framework for interpreting the
characteristic frequencies seen in the 
broad band power spectra from black hole and neutron star binaries.
We use the truncated disc/hot inner flow geometry, and assume that the
hot flow is generically turbulent. Each radius in the hot flow
produces fluctuations, and we further assume that these are damped on
the viscous frequency. Integrating over radii gives broad band
continuum noise power between low and high frequency breaks which are
set by the viscous timescale at the outer and inner edge of the hot
flow, respectively. Lense-Thirring (vertical) precession of the entire
hot flow superimposes the low frequency QPO on this continuum power. 

We test this model on the power spectra seen in the neutron star
systems (atolls) as these have the key advantage that the (upper) kHz
QPO most likely independently tracks the truncation radius. These show
that this model can give a consistent solution, with the truncation
radius decreasing from $20-8~R_g$ while the inner radius of the flow
remains approximately constant at $\sim 4.5~R_g$ i.e. 9.2~km.  We use
this very constrained geometry to {\em predict} the low frequency QPO
from Lense-Thirring precession of the entire hot flow from $r_o$ to
$r_i$. The simplest assumption of a constant surface density in the
hot flow matches the observed QPO frequency to within 25 per cent.
This match can be made even better by considering that the surface
density should become increasingly centrally concentrated as the flow
collapses into an optically thick boundary layer during the spectral
transition. The success of the model opens up the way to use the broad
band power spectra as a diagnostic of accretion flows in strong
gravity.

\end{abstract}


\section{Introduction} \label{sec:introduction}

Black holes and neutron stars have very similar gravitational
potentials as neutron star radii are approximately the size of the
last stable orbit. Thus their accretion flows should be similar
despite the fundamental difference in the nature of the central
object: neutron stars have a solid surface, while black holes do not.
This similarity is seen in the spectral and timing
properties. Both black hole binaries (BHB) and disc accreting neutron
stars (atolls) show a distinct transition between hard spectra seen at
low luminosities (termed the low/hard state in BHB and the island
state in atolls) and much softer spectra seen at high luminosities
(high/soft in BHB, banana branch in atolls).  During the transition,
the properties of the rapid variability also change. This variability
can be approximately described as band limited continuum noise
between a low and high frequency break, with a low frequency
Quasi-Periodic Oscillation (hereafter LF QPO) superimposed. The low
frequency break and LF QPO move to higher frequencies as the
source spectrum softens in both BHB and atolls whereas the high
frequency break is seen to remain approximately constant. Atoll power
spectra also display a pair of kHz QPOs, a feature not (unambiguously)
observed in BHB power spectra. The peak frequency of these QPOs is
seen to increase as the source spectrum softens such that it
correlates with the break frequency and the LF QPO. (see e.g. the reviews
by van der Klis 2005, hereafter vdK05; McClintock \& Remillard 2006,
hereafter MR06; and Done, Gierlinski \& Kubota 2007, hereafter DGK07)

Both spectral and variability behaviour can be qualitatively explained
if the geometrically thin, cool accretion disc is replaced at radius
$r_o$ by a hot inner flow which produces the hard Comptonised
spectrum. As this transition radius decreases, the disc spectrum
increases in luminosity and temperature and more soft seed photons
from the disc illuminate the corona. This increases the Compton
cooling so the hard spectrum softens slightly.  This effect becomes
much stronger when the disc extends far enough down in radii to
overlap with the hot flow, and the Comptonised spectrum softens
dramatically with decreasing radius as the disc approaches the last
stable orbit.  Throughout this evolution, all the timescales for
variability associated with the inner edge of the thin disc
decrease. This broadly explains the correlated spectral-timing behaviour
observed in both BHB and atolls if the low frequency break and LF QPO
are set by $r_o$ (Barret 2001; DGK07).

There have been some quantitative tests of the spectral evolution
predicted by these models as the components which
make up the energy spectra (disc and comptonisation) are well understood
(e.g. Chaty et al 2003; Done \& Gierlinski
2003; Gierlinski, Done \& Page 2008; Cabanac et al 2008).
However, there is no comparable consensus on the 
fundamental components which make up the power
spectra. This is especially evident in the
case of the LF QPO. This is a clear characteristic frequency, most
probably associated with $r_o$, yet it cannot be used to quantitatively
determine $r_o$ until its physical origin is well understood. While there are a
plethora of potential mechanisms in the literature
(e.g. Fragile et al 2001; vdK05; Titarchuk \& Osherovich 1999; 2000), our
recent model of Lense-Thirring (vertical) precession of the hot inner
flow is the first to simultaneously explain both its spectral and
timing properties (Ingram, Done \& Fragile 2009, hereafter IDF09).

Lense-Thirring precession is a relativistic effect whereby a spinning
black hole with its angular momentum misaligned with that of the
binary system produces a torque. This propagates via bending waves
throughout the hot flow, and can make the entire flow within $r_o$
vertically precess as a solid body (though it does not {\em rotate} as
a solid body: orbits are still approximately Keplerian). The
precession frequency is set by the outer radius of the vertically
precessing flow, $r_o$, the dimensionless black hole spin parameter,
$a_*$, and the inner radius of the hot flow, $r_i$, for a given mass,
$M$. The precession frequency of the flow increases as $r_o$ decreases
until $r_o \rightarrow r_i$ when the thin disc extends underneath the
hot flow at all radii, suppressing vertical modes.  IDF09 showed that
this model gives a good match to the observed LF QPO behaviour for BHB
as a class for $r_o$ decreasing $50\to r_i$ (where all radii are in
units of $R_g=GM/c^2$) assuming that $r_i$ is set by the increased
torque from the misaligned flow rather than by the last stable orbit
(Fragile et al 2007, 2009, Fragile 2009).

Here we develop a full model for the power spectrum, encompassing the
broad band noise as well as the LF QPO. The only previous attempt at
such a combined model is Titarchuck \& Osherovich (1999), though the
physical mechanism for modulation of their LF QPO is not clear.
However, the broad band noise components are beginning to be well
understood.  Numerical simulations now give some insight into the
nature of the noise generating process.  Angular momentum transport in
the accretion flow takes place via stresses (a.k.a. `viscosity')
generated by the Magneto-Rotational Instability (MRI: Balbus \& Hawley
1998). This process generates fluctuations in all quantities
(e.g. Krolik \& Hawley 2002). However, the mass accretion rate at any
given radius cannot change faster than the local viscous timescale, so
fluctuations at each radius are damped on this timescale (Lyubarskii
1997; Psaltis \& Norman 2000; also see Titarchuck \& Osherovich 1999;
Misra \& Zdziarski 2008 for a slightly different approach). Coupling
this to the truncated disc/hot flow geometry gives a prediction of
self-similar fluctuation power between timescales corresponding to the
viscous timescale at the inner and outer radii of the hot flow
(Churazov et al 2001; Arevelo \& Uttley 2006). Thus the evolution of
the continuum power spectrum can determine the inner and outer radii
of the flow, and these can be used to {\em predict} the LF QPO
frequency, to compare with that observed.

While we could do this in the BHB systems, the atolls give additional
constraints as the spin of the neutron star is often independently
known from burst oscillations (e.g. Strohmayer, Markwardt, \& Kuulkers
2008; Piro \& Bildsten 2005). Neutron star power spectra also contain
the upper
and lower kHz QPOs (e.g. vdK05) together with an additional high
frequency noise component (Sunyaev \& Revnivtsev 2000). The upper kHz
QPO (ukHz QPO) is most likely the Keplerian frequency at the
truncation radius $r_o$ (e.g. van der Klis et al 1996; Stella \&
Vietri 1998; Schnittman 2005) and the narrowness of the feature means
that this gives an unambiguous determination of $r_o$. This
identification independently constrains a key parameter of the LF QPO
model. Hence here we use the atolls to outline a self-consistent model
for {\em all} the observed components in the power spectrum.

\section{The origin of the broad band power spectrum} \label{sec:pow}

We choose atoll sources with multiple observations showing the power
spectral evolution so as to test the model over a wide range of $r_o$.
We consider only low spin systems ($a_*<0.3$), because higher spins
lead to an equatorial bulge of the neutron star which distorts
space-time from being well described by the Kerr metric (Miller et al
1998). This leads us to pick the atoll systems 4U 1728-34 and 4U
0614+09 (van Straaten et al 2002), both of which have spin $a_*\sim
0.2$ and (assumed) mass $M\sim 1.4M_{\sun}$.

\begin{figure}
\centering$
\begin{array}{c}
\leavevmode  \epsfxsize=6.5cm \epsfbox{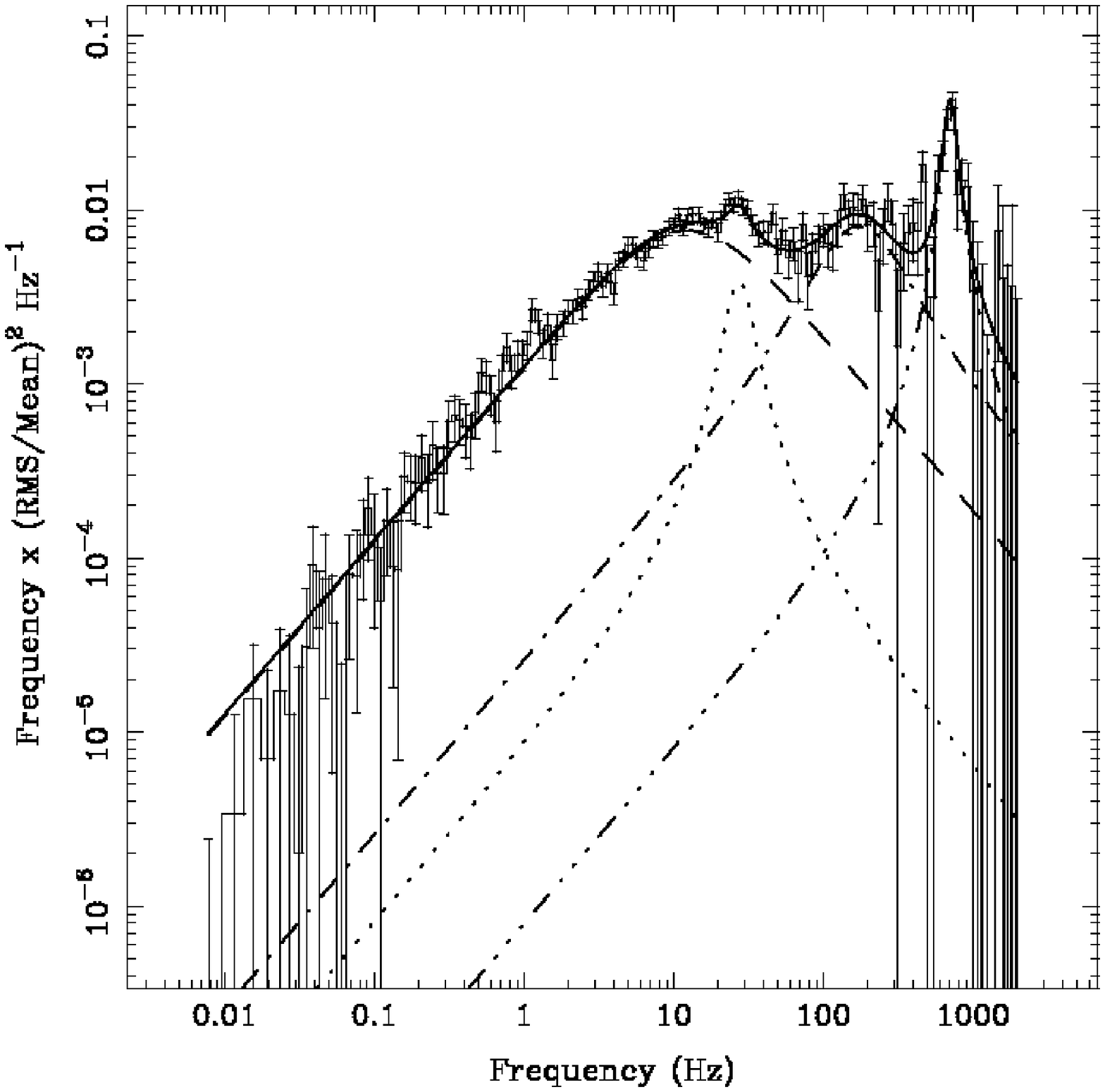}\\
\leavevmode  \epsfxsize=6.5cm \epsfbox{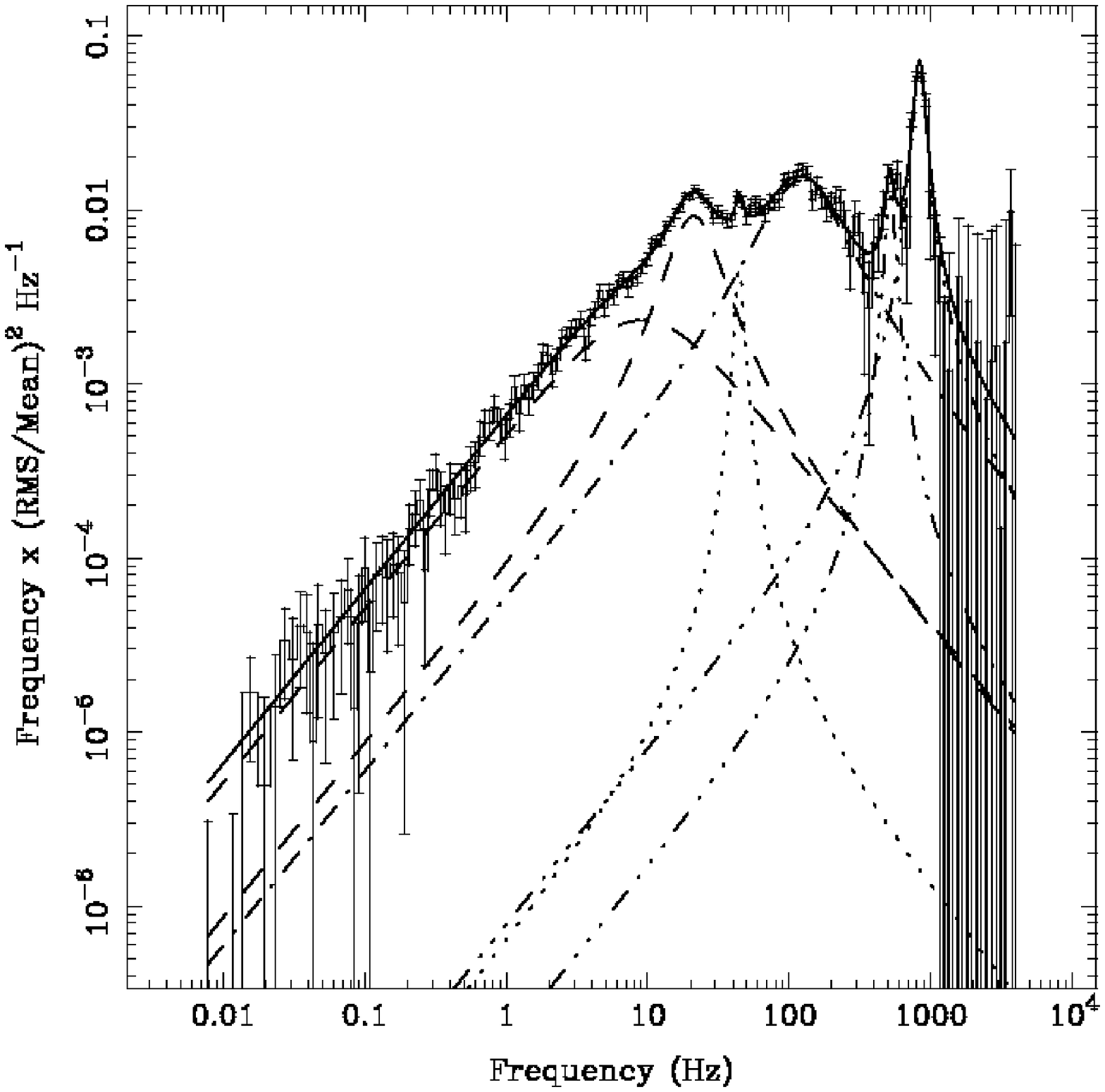}
\end{array}$
\caption{Power spectra and fit functions for 4U 1728-34 (top) and
4U 0614+09 (bottom), reproduced with the permission of van Straaten et
al (2002) and the AAS. Lorentzians represent the following components:
the lower break $L_b$ (dashed), the LF QPO (dotted), the high
frequency break 
$L_h$ (dot-dashed) and the kHz QPOs (triple dot-dashed). When there are
two dashed lines present, as in the bottom panel, we will refer
to the left hand one as $L_{b2}$ and the right hand one as $L_{VLF}$
with one assumed to be a continuation of $L_b$.}
\label{fig:neutron}
\end{figure}

Typical power spectra of 4U 1728-34 and 4U 0614+09 are shown in the
top and bottom panels respectively of Figure \ref{fig:neutron}. We see
that the QPOs described above are superimposed on a complex, broad
band noise continuum. This can be approximately modelled by a
twice broken power law, following $P(\nu)\propto \nu^{0}$ below the
lower break frequency $\nu_b$, $P(\nu)\propto \nu^{-2}$ above the higher
break frequency $\nu_h$ and $P(\nu)\propto \nu^{-1}$ between (i.e a flat
top in $\nu P(\nu)$). However, multiple Lorentzians give a much better
description of the broad band noise (Belloni, Psaltis \& van der Klis
2002; Fig 1).  Each of these has a characteristic frequency, $\nu_c$, and
width, $\Delta \nu$, which can be combined together into a quality factor
$Q=\nu_c/\Delta \nu$. The lowest frequency component, $L_b$, is generally
a zero centred Lorentzian, so it peaks in $\nu P(\nu)$ at $\Delta \nu$,
producing the low frequency `break' in the flat top noise. This break
frequency correlates with the LF QPO and kHz QPOs (Wijnands \& van der
Klis 1999; Klein-Wolt \& van der Klis 2008; Psaltis, Belloni \& van
der Klis 1999) whereas the high frequency `break' (sometimes referred to as
the hectohertz QPO) varies much less (e.g. vdK05; DGK07).

\subsection{Outer radius}

This behaviour of the high and low frequency breaks can be
qualitatively explained in the truncated disc/hot inner flow model.
The inner radius of the flow remains constant at the neutron
star radius, so giving the constant high frequency power, while the
outer radius sweeps inwards, leading to the progressive loss of low
frequency components (Gierlinski, Nikolajuk \& Czerny 2008).
Quantitatively this can be modelled by each radius generating noise
power as a zero centred Lorentzian with width $\Delta \nu=\nu_{visc}$.
The viscous frequency $\nu_{visc} = 1.5 \alpha (h/r)^2\nu_\phi$,
where $\alpha$ is the Shakura-Sunyaev viscosity parameter,
$h/r$ is the disc semi-thickness and $\nu_\phi$ is the rotational
frequency of fluid particles within the flow. However, none of these
are necessarily straightforward to define. MRI simulations of black
hole accretion flows show that $\alpha$ and $h/r$ vary with radius
(e.g. Fragile et al 2007, 2009). Additionally, $h/r$ should change
during state transitions as
the hot inner flow collapses.  In neutron stars especially, this
collapse marks the transition from the hard X-ray emission region
being an extended optically thin boundary layer which merges smoothly
onto the hot inner flow, to a much more compact boundary layer.
As well as the impact of such a transition on $h/r$, the viscosity
mechanism in the boundary layer may well be very different to that of
the standard MRI, and the azimuthal velocity field is dominated by
that of the star rather than being Keplarian. 

\begin{figure}
\centering
\leavevmode  \epsfxsize=6.5cm \epsfbox{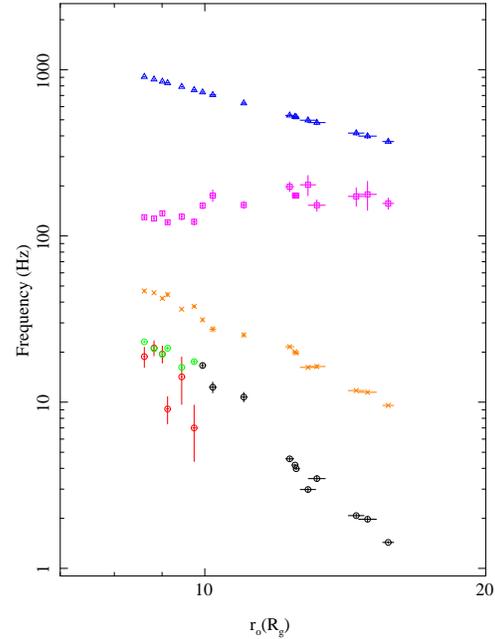}
\caption{
Plot of characteristic frequencies plotted against $r_o$ as
inferred from the assumption $\nu_{ukHz}=\nu_k(r_o)$.
The blue triangular points represent $\nu_{ukHz}$ and the
square magenta points represent $\nu_h$. The orange crossed
points represent the LF QPO frequencies and the
circular points the low frequency break.
The black points are for power spectra where there is no ambiguity over
what the break frequency is whereas the red points are for $\nu_b=\nu_{b2}$
and the green points for $\nu_b=\nu_{VLF}$.}
\label{fig:alpha}
\end{figure}

This makes neutron stars somewhat more complex than black
holes. However, their saving grace is that we can use their additional
kHz QPOs to independently determine $r_o$ assuming that
$\nu_{ukHz}=\nu_k(r_o) = c/[2\pi R_g (r_o^{3/2}+a_*)]$ (it should be
safe to assume $\nu_\phi$ at the inner edge of the \textit{disc} to be
Keplerian). The blue triangular points in Figure \ref{fig:alpha} show
that this requires $r_o$ to decrease from $20-8 R_g$, consistent with
the expected change in radius from the spectral softening seen from
the island state to the lower banana branch (Barret 2001). 

The square magenta points in Figure \ref{fig:alpha} show the high
frequency break (hectohertz) frequency, which remains approximately constant as
discussed earlier and the crossed orange points show the LF QPO
frequency. This correlates with the low frequency break (e.g. Wijnands
\& van der Klis 1999), which is represented by the circular points. Of
these, the black points represent data where $\nu_b$ is unambiguously
identified in the power spectra. However, this becomes difficult at
the highest ukHz QPO frequencies (i.e. smallest radii) as there is an
additional component observed in the low frequency power spectrum
e.g. the lower panel of Figure~\ref{fig:neutron}, where two low
frequency Lorentzians are required. It is not immediately clear which
one of these corresponds to $\nu_b$ e.g. van Straaten et al (2002)
refer to the lowest frequency Lorentzian as $L_b$ and call the other
$L_{VLF}$ while Altamirano et al (2008) put $L_b$ on the right and
term the other $L_{b2}$. Here we only use $L_b$ where this break is
unambiguously determined by the data. Where there are two competing
low frequency components we refer to the lowest frequency one as
$L_{b2}$ and the other as $L_{VLF}$. The green points in Figure
\ref{fig:alpha} represent $\nu_{VLF}$ whereas the red points represent
$\nu_{b2}$.  The green points connect smoothly onto the black points
where $\nu_b$ is unambiguously determined, while the red points do
not.  Thus it seems most likely that the higher of the two low
frequency components represents the continuation of the break
frequency determined by $r_o$.

Of these 6 points with a split break frequency, 4 are from observations
of 4U 1728-34 and 2 from 4U 0614+09. If we analyse the colour-colour
diagram of 4U 1728-34 (Di Salvo et al 2000), we see that these 4
observations (9-12 of 19) occur just before the transition between the
island state and the banana branch. Intriguingly, the geometry inferred
from models of the spectral evolution require an {\em overlap} between
the hot flow and truncated disc close to the transition. The splitting
of the break frequency then has an obvious interpretation with the outer
radius of the hot flow being larger than the inner radius of the thin
disc.  The hot flow in this overlap region will have smaller scale
fluctuations, as the size scale of the magnetic field is limited by the
thin disc in the mid plane. Thus $\nu_{b2}$ can be interpreted as the
viscous frequency at the edge of the corona with $\nu_{VLF}$ being the
viscous frequency at the truncation radius.

\subsection{Inner radius}

We assume that the lower break frequency, $\nu_b$ is identified with
$\nu_{visc}(r_o)$, and we use the independent constraints on $r_o$
from the ukHz QPO above to track out the unknown variation in
$\nu_{visc}$ ($\propto \alpha (h/r)^2\nu_\phi(r)$). We parameterise
this as a power law, so that $\nu_b=\nu_{visc}(r_o)= A
r_o^{-\gamma}$. We then use the best fit values of $A$ and $\gamma$
derived from the low frequency break to determine $r_i=
[(A \nu_h)]^{1/\gamma}$ assuming that the high frequency break in the
noise power (hectohertz component) is the viscous frequency at $r_i$.

\begin{figure}
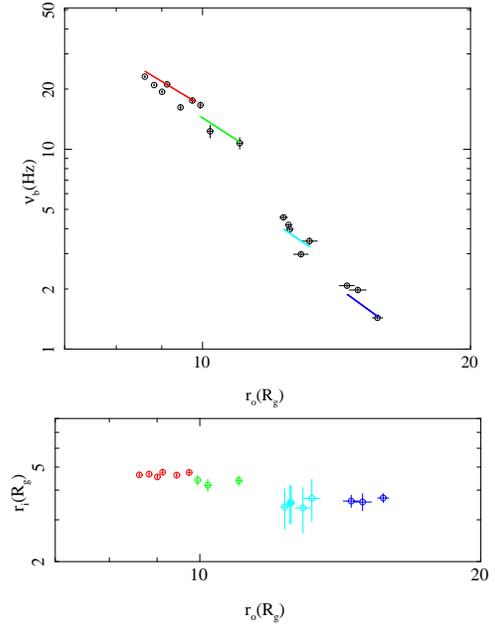

\centering$
\begin{array}{c}
\leavevmode  \epsfxsize=6.2cm \epsfbox{4break.ps} \\
\leavevmode  \epsfxsize=6.5cm \epsfbox{4rinner.ps}
\end{array}$
\caption{
\textit{Top panel:} Break frequency plotted against truncation radius, $r_o$,
with four separate power law fits: $\gamma=3.25$ (blue), $3.02$ (magenta),
$2.88$ (green) and $2.69$ (red). This treatment assumes that the
viscous frequency is given by a power law, the index of which becomes
less negative as $r_o$ reduces. $\nu_b$ is then $\nu_{visc}(r_o)$
and $r_i$ is the value of $r$ that gives $\nu_{visc}(r)=\nu_h$.
\textit{Bottom panel:} Inferred values for $r_i$ plotted against $r_o$.}
\label{fig:rinner}
\end{figure}

However, as discussed in the previous section, we do not necessarily
expect this power law representation of $\nu_{visc}(r)$ to stay
constant as the truncation radius sweeps in and the source spectrum
softens due to the collapse of the more extended hot flow into the
boundary layer, with its potentially very different viscosity and
azimuthal velocity. Instead we split the radial range in $r_o$ into
four groups of points, each described by a different best fit power
law. The top panel of Figure \ref{fig:rinner} shows this best fit
power law relation for each group of points, with a clear change in
both slope and normalisation as the 
truncation radius moves inwards. Quantatively, the inferred value of
$\gamma$ moves from $3.25$ (blue), $3.02$ (magenta), $2.88$ (green) to
$2.69$ (red).  We can now use our moving power law representation in
order to extrapolate values for $r_i= [(A\nu_h)]^{1/\gamma}$ taking
care to use the correct values of $\gamma$ and $A$ for a given value
of $\nu_h$.

The lower plot of Fig.~\ref{fig:rinner} shows the derived values for
$r_i$ with error bars including the systematic error in determining
the best fit values of $A$ and $\gamma$. We infer from this that the
radius of the neutron star lies at $r_i \approx 4.5 \pm 0.04 \approx
9.2 \pm 0.1$~km. This would mean that the neutron star is slightly
smaller than its own last stable orbit ($5.3~R_g$ for $a_*=0.2$),
indicating a soft equation of state, but we caution that the exact
value depends on the accuracy of our assumed power law representation
of the viscous frequency with radius. Any more complex form will
extrapolate to a different inner radius, and the value of this radius
may also be affected by time dilation.  Nonetheless, the remarkable
constancy of the derived inner radius gives some confidence in our
approach, and the value of 9.2~km is very close to the `canonical'
assumption of 10~km for a $1.4M_\odot$ neutron star.

\section{Testing Lense-Thirring in atolls} \label{sec:lt}

\begin{figure}
\centering
\leavevmode  \epsfxsize=6.5cm \epsfbox{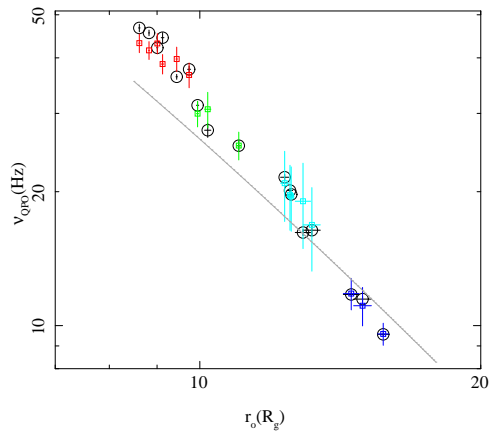}
\caption{LF QPO frequency plotted against truncation radius
(black circles). The grey line tracks Lense-Thirring
precession frequency of the inner flow with $r_i=4.5$ and
$\zeta=0$. The blue, magenta, green and red squares are for
$\zeta=-0.7$, $-0.3$, $0.6$ and $2.7$ respectively
and use the $r_i$ values from the bottom plot of Figure \ref{fig:rinner}.}
\label{fig:lf}
\end{figure}

Now we have both the inner and outer radius for the hot flow, we can
directly calculate the predicted Lense-Thirring precession
frequency. However, there is one additional free parameter which is
the mass distribution in the hot flow, which can be parameterised by
$\zeta$, the radial dependence of the surface density,
$\Sigma=\Sigma_i(r/r_i)^{-\zeta}$ (see IDF09 and Fragile et al 2007).
The LF QPO frequency is then {\em predicted} to be 
\begin{equation}
\nu_{prec} = \frac{(5-2\zeta)}{\pi(1+2\zeta)}
\frac{a_*[1-(r_i/r_o)^{1/2+\zeta}]}
{r_o^{5/2-\zeta}r_i^{1/2+\zeta}
[1-(r_i/r_o)^{5/2-\zeta}]}\frac{c}{R_g}
\label{eqn:tprec}
\end{equation}

Simulation data for black holes shows $\zeta\sim 0$ (e.g. Fragile et
al 2007) but neutron stars have a solid surface which could give a
rather different situation where the flow is increasingly concentrated
on the neutron star surface as the accretion rate increases.
Nonetheless, assuming $\zeta= 0$, and taking $r_i$ fixed at $4.5$
(see previous section) gives quite a good fit (grey line) 
to the observed LF QPO (black circles) as shown in Fig \ref{fig:lf}.

The fit can be made even better by allowing $\zeta$ to vary.  As $r_o$
decreases the expectation is that the flow goes from being similar to
the BH case, to being more and more concentrated in the boundary layer
i.e. we expect an increase in $\zeta$ as the dense boundary layer
begins to dominate the surface density of the flow. Such an increase
in the surface density profile is also implied by the change in
viscous frequency implied from the previous section, since surface
density is inversely proportional to the radial velocity
$v_r=R\nu_{visc}$.  We fit our Lense-Thirring model to the four
different sets of points from before and obtain excellent agreement
with observation if $\zeta$ takes the values $-0.7$ (blue), $-0.3$
(magenta) $0.6$ (green) and $2.7$ (red) i.e.  $\zeta$ increases with
decreasing $r_o$ as expected.  However, a quantitative understanding of
how these parameters should interact in neutron stars is a very
difficult goal as the boundary conditions associated with accreting
neutron stars are so poorly understood.

\section{Conclusions}

We show that the broadband continuum noise power and LF QPO seen in
atolls and BHB can be self-consistently explained in the {\em same}
truncated disc/hot inner flow model which describes their spectral
evolution. We test this on the atoll systems, as these have strong kHz
QPOs which most probably pick out the truncation radius of the thin
disc, $r_o$, so this key parameter is known independently. Using the
standard assumption that the upper of the two kHz QPOs marks the
Keplerian frequency gives that $r_o$ decreases from $20-8~R_g$ during
the marked spectral transition seen in atolls from the hard (island)
state to soft (banana branch) spectra.

The low frequency break seen in the noise power is then consistent
with being the viscous timescale of the hot flow at $r_o$.  All
smaller radii in the hot flow contribute to the noise power, giving
the broad band continuum power spectrum. The highest frequency noise
component marks the viscous timescale at the inner edge of the hot
flow, $r_i$.  We use our parameterisation of $\nu_{visc}$ to calculate
$r_i$ and find that this remains remarkably constant at $r_i\sim
4.5\equiv$ 9.2~km for a $1.4M_\odot$ neutron star.

The truncated disc model also gives a physical interpretation for the
observed `splitting' of the lowest frequency noise component seen
close to the spectral transition. At this point the spectral models
predict that the disc overlaps the hot flow, so there is a component
which tracks turbulence in the hot flow {\em within} the disc inner
radius, and another component which tracks the true outer edge of the
hot flow which extends over the disc.

With all of the parameters of the truncated disc geometry constrained,
we are then able to test the Lense-Thirring precession model for the
LF QPO presented in IDF09. This gives a fairly good match to the data
at large truncation radii, but increasingly underestimates the QPO
frequency as $r_o$ decreases. Nonetheless, it still only 25 per cent
too low even at the smallest $r_o$. However, there is still one
additional free parameter which is the radial dependence of the
surface density of the hot flow. Allowing this to change so that the
flow becomes increasingly concentrated towards $r_i$ as the truncation
rate decreases, as expected from the collapse of a hot flow into the
boundary layer, gives an excellent match to the data.  However, we
caution that the expected evolution of the surface density is not well
understood quantitatively for neutron stars.

It must also be noted that considering the whole flow to precess removes
a previous objection to Lense-Thirring precession as the origin of the
LF QPO. If the LF QPO is produced by Lense-Thirring at $r_o$ then this
implies the moment of inertia of the neutron star is too large (Markovic
and Lamb 1998). Instead, in our model the LF QPO is produced at some mass
weighted radius between $r_o$ and $r_i$ with the weight increasingly
towards $r_i$ for softer spectra (higher frequencies). Thus for the
lowest values of $r_o\sim 8.5$, the LF QPO is predominantly produced
by material at $r_i=4.5$ rather than at $r_o$, so the moment of
inertia is correspondingly reduced. 

Overall, we present a  model of the power spectrum in which
both broad band continuum and LF QPO components are interpreted
physically. This forms a framework in which the characteristic
frequencies in the power spectrum can be used as a diagnostic of the
properties of the accretion flow in strong gravity.

\section{Acknowledgements}

AI acknowledges the support of an STFC studentship. AI and CD acknowledge
useful comments from Chris Fragile, Cole Miller, Didier Barret and the
referee Lev Titarchuk.


\label{lastpage}

\end{document}